\documentclass[showpacs,amsmath,amssymb,aps]{revtex4}
\usepackage{graphicx}
\newcommand{\abs}[1]{\vert#1\vert}
\renewcommand{\vec}{\bf}
\newcommand{\vecn}{\bf\it}
\newcommand{\bnabla}{\boldsymbol\nabla}
\DeclareMathOperator{\rmi}{i}
\DeclareMathOperator{\rmd}{d}
\DeclareMathOperator{\rme}{e}
\newcommand{\Hessian}{\mathcal{H}^{{\vecn r}^*}}
\begin{document}
\bibliographystyle{apsrev}
\title{Spectral properties of spherically confined dusty plasma crystals}
\author{C. Henning$^1$}
\author{H. K\"ahlert$^1$}
\author{P. Ludwig$^1$}
\author{A. Melzer$^2$}
\author{M. Bonitz$^1$}
\email{bonitz@physik.uni-kiel.de}
\affiliation{$ˆ1$ Institut f\"ur Theoretische Physik und Astrophysik, Christian-Albrechts-Universit\"at zu Kiel, D-24098 Kiel, Germany}
\affiliation{$ˆ2$ Institut f\"ur Physik, Universit\"{a}t Greifswald, D-17489 Greifswald, Germany}
\begin{abstract}
A combined theoretical and experimental analysis of the normal modes of three-dimensional spherially confined Yukawa clusters is presented. Particular attention is paid to the breathing mode and the existence of multiple monopole oscillations in Yukawa systems. Finally, the influence of dissipation on the mode spectrum is investigated.
\end{abstract}
\date{\today}
\maketitle
\section{Introduction}
Spherically confined dust crystals have recently attracted large attention, see e.g. \cite{morfill07} and \cite{bonitz_classical_2008} for an overview. After the exploration of the ground state structure \cite{Hasse91,Ludwig05,bonitz_strutural_2006,Apolinario07} and of metastable states and their probability  \cite{Block08,Kaeding08,Kaehlert08} a question of particular interest is the dynamical excitation spectrum. 
The normal modes are a key property describing the response of a finite system to external excitation and also the melting behavior, e.g. \cite{schweigert_spectral_1995,dubin_normal_1996}. Particularly interesting is the situation in dusty plasmas where the normal mode spectrum of two-dimensional crystals could be analyzed experimentally, e.g. \cite{melzer_normal_2001,Melzer03}. Here, we extend this analysis to three-dimensional spherical Yukwa crystals (Yukawa balls) and present experimental results. 
From the theory side, we recall the determination of the normal modes of confined systems. We concentrate on one of the key modes -- the breathing mode (BM) describing uniform expansion and contraction of the whole system. Recently it was shown \cite{henning_existence_2008} that a ``true'' (i.e. uniform) BM does not exist in Yukawa systems. Instead, there may exist several similar modes which, however, all deviate either from uniform or radial motion of the particles. The brief derivation of Ref. \cite{henning_existence_2008} is reproduced here in a more detailed way. One of the important consequences -- the possible existence of multiple monopole modes -- is demonstrated. Finally, we take into account the effect of dissipation on the normal mode spectrum which is crucial for dusty plasmas.
%
\section{Normal modes of finite clusters}\label{Normal_modes_of_finite_clusters}
The crystals are characterized by a $3$-dimensional, classical system of $N$ identical particles harmonically confined by the potential $\phi(r)=m \omega_0^2 r^2/2 $ and interacting with a Yukawa-type pair interaction $v(r)=q^2 \exp(-\kappa r)/r$. The hamiltonian is then given by ($\sum'$ indicates no summation over equal indices)
\begin{equation}\label{eq:hamiltonian}
	H=\sum_{i=1}^N\frac{{\vec p}_i^2}{2m} +\underbrace{\sum_{i=1}^N \phi(\abs{{\vec r}_i})
	+\frac{1}{2}\sideset{}{'}{\sum}_{i,j=1}^N v(\abs{{\vec r}_{ij}})}_{U({\vecn r}\,\in\,\mathbb{R}^{3N})}.
\end{equation}
To investigate the spectral properties of dusty plasma crystals at low temperatures we consider small excitations from a ground state or metastable state ${\vecn r}^*=\bigl({\vec r}^*_1,{\vec r}^*_2\ldots{\vec r}^*_N\bigr)\,\in\mathbb{R}^{3N}$, which is a minimum of the potential energy and thus fulfills the equations
\begin{equation}\label{eq:equilibrium_positions}
	{\vec 0}=\nabla_i U({\vec r})\vert_{{\vec r}={\vec r}^*}=\frac{\phi'(\abs{{\vec r}^*_i})}{\abs{{\vec r}^*_i}}{\vec r}^*_i+\sideset{}{'}\sum_{\substack{l=1}}^{N}\frac{v'(\abs{{\vec r}^*_{il}})}{\abs{{\vec r}^*_{il}}}{\vec r}^*_{il}\hfill\forall i\leq N.\hspace{1cm}
\end{equation}
The small excitation is a time-dependent function ${\vecn r}(t)$ in the configuration space with $\abs{{\vecn r}(t)-{\vecn r}^*}\ll1$. Due to the small amplitude of the oscillation the potential can be approximated harmonically by
\begin{equation}\label{eq:harmonic_approximation}
	U({\vecn r})\approx U({\vecn r}^*)+\frac{1}{2}({\vecn r}-{\vecn r}^*)^T\Hessian({\vecn r}-{\vecn r}^*)
\end{equation}
where we used (\ref{eq:equilibrium_positions}) and the definition of the Hessian matrix $\Hessian=\bnabla\bnabla^T U({\vecn r})\vert_{{\vecn r}={\vecn r}^*}$. Because $\Hessian$ is a real symmetric $3N\times3N$-matrix and positive semidefinite as well, its eigenvalue problem \cite{henning_existence_2008}
\begin{equation}\label{eq:eigenvalue_problem}
	\lambda m {\vecn r}=\Hessian{\vecn r}
\end{equation}
defines $3N$ eigenvalues $\lambda_j\geq0$ and $3N$ linearly independent eigenvectors ${{\vecn r}_j}$, which form a basis in the configuration space and may be conveniently chosen orthonormal. Within this basis the excitation can be expanded
\begin{equation}\label{eq:mode_expansion}
	{\vecn r}(t)={\vecn r}^*+\sum_{j=1}^{3N} c_j(t){\vecn r}_j,
\end{equation}
so that the time-dependence is fully determined by the coefficients $c_j(t)$ -- the normal coordinates. These normal coordinates obey equations of motion, which follow from the hamiltonian (\ref{eq:hamiltonian})
\begin{equation}\label{eq:newton}
	\begin{split}
		{\vec 0}
			&\overset{(\ref{eq:hamiltonian})}{=}m\ddot{{\vecn r}}+\bnabla U({\vecn r})
			\overset{(\ref{eq:harmonic_approximation})}{=}m\ddot{{\vecn r}}+\Hessian({\vecn r}-{\vecn r}^*)\\
			&\overset{(\ref{eq:mode_expansion})}{=}m\sum_{j=1}^{3N} \ddot{c}_j(t){\vecn r}_j+\sum_{j=1}^{3N} c_j(t)\Hessian{\vecn r}_j
			\overset{(\ref{eq:eigenvalue_problem})}{=}m\sum_{j=1}^{3N}\left[\ddot{c}_j(t)+\lambda_j c_j(t)\right]{\vecn r}_j.
	\end{split}
\end{equation}
Because the eigenvectors ${{\vecn r}_j}$ are linearly independent each normal coordinate $c_j(t)$ fulfills $0=\ddot{c}_j(t)+\lambda_j c_j(t)$ with the solution
\begin{equation}
	c_j(t)=A_j\cos(\sqrt{\lambda_j}t+B_j)\hfill\forall j\leq 3N,\hspace{1cm}
\end{equation}
in which the constants $A_j$ and $B_j$ have to be determined from the initial conditions ${\vecn r}(0),\dot{\vecn r}(0)$ of the excitation. Thus in general, the excitation relative to the ground state, ${\vecn r}(t)-{\vecn r}^*$, is a superposition of oscillating motions $A_j\cos(\sqrt{\lambda_j}t+B_j){\vecn r}_j$, called normal modes. Each such normal mode describes a collective motion of all particles oscillating with the same frequency $\omega_j=\sqrt{\lambda_j}$ -- the normal frequency. The respective displacements of the particles are given by the eigenvector ${\vecn r}_j$. Consequently, in order to investigate the spectral properties of the dusty plasma crystals we have to investigate the normal modes, and hence have to consider the eigenvalue problem (\ref{eq:eigenvalue_problem}) of the Hessian $\Hessian$.

\begin{figure}
	\begin{center}
		\includegraphics[width=\textwidth,clip=true]{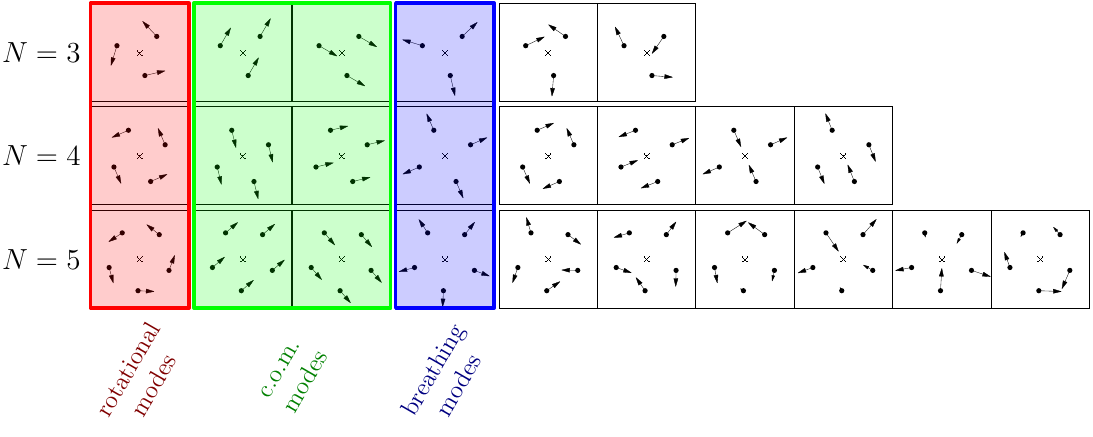}
	\end{center}
	\caption{(Color online) All normal modes of two-dimensional harmonically confined Coulomb systems with $N=3,4,5$ particles. The dots picture the particles within a ground state configuration, and the arrows show the direction of the oscillatory motion. The $N$-independent modes, i.e., the rotational modes, the modes of center of mass oscillation (c.o.m.), and the breathing modes are highlighted.}
\label{fig:normal_modes}
\end{figure}

For harmonically confined Coulomb systems ($\kappa=0$) detailed theoretical studies have been performed for $d=1,2,3$ dimensions, see \cite{schweigert_spectral_1995,dubin_normal_1996} and references therein. It was shown that there exist three (partially degenerate) normal modes, which are independent of the particle number $N$:
\begin{enumerate}
	\item The $d(d-1)/2$ rotational modes with $\lambda=0$, which correspond to a rotation of the whole system and reflect the axial symmetries of the confinement.
	\item The $d$ center of mass oscillation modes with $\lambda=\omega_0^2$ expressing that the center of mass motion is independent of the interparticle forces.
	\item The breathing mode (BM) with $\lambda=3\,\omega_0^2$, which describes a uniform radial expansion and contraction of all particles.
\end{enumerate}
The existence of these three modes is illustrated for the two-dimensional system with $N=3,4,5$ particles in figure \ref{fig:normal_modes}, where all modes of these systems corresponding to the ground state configuration are shown.

\section{Normal modes in Yukawa systems}
\begin{figure}
	\begin{center}
		\includegraphics[width=10cm,clip=true]{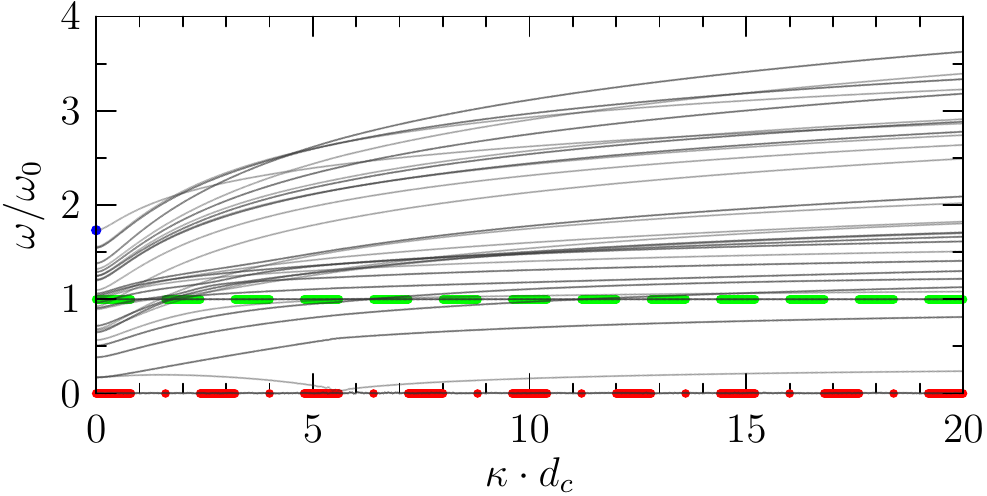}
	\end{center}
	\caption{(Color online) $\kappa$-dependence of the normal modes of a harmonically confined Yukawa system with $N=16$ particles vs. screening parameter. The rotational modes, the center of mass oscillations and the breathing mode are highlighted dash-dotted (red), dashed (green), and (blue) dot, respectively.}
\label{fig:modes_with_screening}
\end{figure}

For harmonically confined Yukawa systems the interaction potential and thus the Hessian $\Hessian$ depend on a screening parameter $\kappa$. Hence, in general, the normal modes will depend on this parameter as well. Within figure \ref{fig:modes_with_screening} this situation is shown.
There, the frequencies of all normal modes are shown for the ground state of harmonically confined Yukawa systems with $N=16$ particles and a screening parameter ranging from $\kappa=0$ to $\kappa=20 d_c^{-1}$ \cite{endnote}. Moreover, the figure shows, that there are two mode frequencies independent from screening -- corresponding to the rotational modes and the center of mass oscillation modes. These modes exist also in a Yukawa system (and have $\kappa$-independent frequencies) due to the symmetries of the systems \cite{partoens_classial_1997}. However, in general this is not the case for the breathing mode, which has no $\kappa$-independent frequency. Actually, the general existence of this mode is a special property of Coulomb-like interactions not featured in Yukawa systems \cite{henning_existence_2008}.

The existence condition of the BM becomes clear by analyzing the eigenvalue problem (\ref{eq:eigenvalue_problem}). The standard procedure to evaluate this equation is to find the roots of its characteristic polynomial yielding all eigenvalues and -vectors and to examine whether there exists an eigenvector ${\vecn r}_{\rm BM}$ corresponding to a BM, i.e., describing a uniform and radial motion ${\vecn r}_{\rm BM} \propto {\vecn r}^*$. However, the degree of this equation is $d\!\cdot\! N$ which prohibits an analytical calculation of the eigenvalues and -vectors and, consequently, prohibits general statements about the existence of the BM. Therefore, we apply a different approach -- the direct mode approach -- which focuses on just one mode: within the eigenvalue equation we directly use the eigenvector of the breathing mode, which thus leads to the existence condition of this mode.

Additionally, we evaluate (\ref{eq:eigenvalue_problem}) within its particle components $i\leq N$, because the potentials $\phi$ and $v$ are given for these, and obtain
\begin{equation}\label{eq:eigenvalue_problem_components}
	\lambda m {\vec r}_i=\sum_{l=1}^N\Hessian_{il}{\vec r}_l=\sum_{l=1}^N\nabla_i\nabla^T_l U({\vecn r})\vert_{{\vecn r}={\vecn r}^*}{\vec r}_l.
\end{equation}
Using the isotropy of $\phi$ and the distance dependence of $v$ one obtains for each component of the Hessian
\begin{equation}
\begin{split}
	&\nabla_i\nabla^T_l U({\vecn r})\vert_{{\vecn r}={\vecn r}^*}=\\
		&\quad\biggl[\frac{\phi'(\abs{{\vec r}_l^*})}{\abs{{\bf r}_l^*}}{\mathcal I}_3
			+\frac{{\vec r}_l^*{{\vec r}_l^*}^T}{\abs{{\vec r}_l^*}^3}
		\Bigl(\abs{{\vec r}_l^*}\phi''(\abs{{\vec r}_l^*})-\phi'(\abs{{\vec r}_l^*})\Bigr)\biggr]\delta_{il}\\
		&\qquad+\sideset{}{'}\sum_{\substack{k=1}}^N
		\biggl[\frac{v'(\abs{{\vec r}_{lk}^*})}{\abs{{\vec r}_{lk}^*}}{\mathcal I}_3+\frac{{\vec r}_{lk}^*{{\vec r}_{lk}^*}^T}{\abs{{\vec r}_{lk}^*}^3}
		\Bigl(\abs{{\vec r}_{lk}^*}v''(\abs{{\vec r}_{lk}^*})-v'(\abs{{\vec r}_{lk}^*})\Bigr)\biggr]\Bigl(\delta_{il}-\delta_{ik}\Bigr),
	\end{split}
\end{equation}
wherein ${\mathcal I}_3$ denotes the three-dimensional identity matrix. Thus (\ref{eq:eigenvalue_problem_components}) reduces to
\begin{equation}
\begin{split}
		\lambda m {\vec r}_i
		&=\frac{\phi'(\abs{{\vec r}_i^*})}{\abs{{\vec r}_i^*}}{\vec r}_i
			+\frac{({\vec r}_i^*\cdot{\vec r}_i){\vec r}_i^*}{\abs{{\vec r}_i^*}^3}\Bigl(\abs{{\vec r}_i^*}\phi''(\abs{{\vec r}_i^*})-\phi'(\abs{{\vec r}_i^*})\Bigr)\\
		&\quad+\sideset{}{'}\sum_{\substack{l=1}}^N
		\biggl[\frac{v'(\abs{{\vec r}_{il}^*})}{\abs{{\vec r}_{il}^*}}{\vec r}_{il}+\frac{({\vec r}_{il}^*\cdot{\vec r}_{il}){\vec r}_{il}^*}{\abs{{\vec r}_{il}^*}^3}
		\Bigl(\abs{{\vec r}_{il}^*}v''(\abs{{\vec r}_{il}^*})-v'(\abs{{\vec r}_{il}^*})\Bigr)\biggr].
	\end{split}
\end{equation}
Now we use the aforementioned direct mode approach, i.e., we set the eigenvector ${\vecn r}$ equal to the unit eigenvector of the BM ${\vecn r}_{\rm BM}=|{\vecn r}^*|^{-1}{\vecn r}^*$, which consequently leads to the existence conditions of this mode
\begin{equation}\label{existing_condition}
	\lambda_{\rm BM} m {\bf r}_i^*=\phi''(\abs{{\bf r}_i^*}){\bf r}_i^*+\sideset{}{'}\sum_{\substack{l=1}}^N v''(\abs{{\bf r}_{il}^*}){\bf r}_{il}^*\hfill\forall i\leq N.\hspace{1cm}
\end{equation}
These conditions show \cite{henning_existence_2008}, that a breathing mode only exists in harmonically confined systems with interaction potentials of the form $1/r^\gamma$ $(\gamma\in\mathbb{R}_{\neq0})$ or $\log(r)$, or in some highly symmetric systems.

The breathing mode and its existence is closely related to the monopole oscillation (MO) \cite{dubin_normal_1996}, which in strongly correlated systems is associated with the oscillation of the mean square radius $R^{\hspace{-0.5pt}2}(t)=N^{-1}\sum_{i=1}^N {\vec r}_i(t)^2=N^{-1}\,{\vecn r}(t)^2$ \cite{schweigert_spectral_1995,partoens_classial_1997,melzer_normal_2001}. This MO is particulary important since it can be easily excited selectively by variation of the trap frequency $\omega_0$, e.g., by a rapid change of the frequency $\omega_0$ of the equilibrated system corresponding to the excitation given by ${\vecn r}(0)=(1+\xi){\vecn r}^*$ and $\dot{\vecn r}(0)={\vec 0}$.

\begin{figure}
	\begin{center}
		\includegraphics[width=7cm,clip=true]{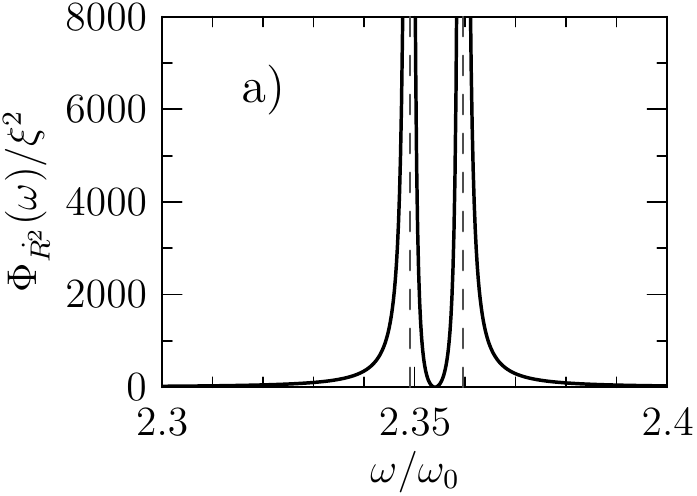}
		\includegraphics[width=5cm,clip=true]{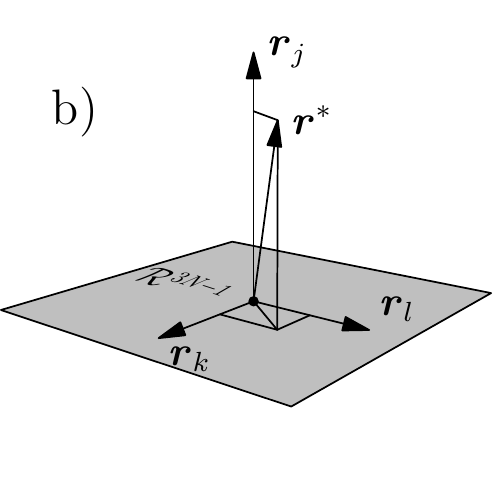}
	\end{center}
	\caption{a) Part of the spectrum $\Phi_{\dot{R^{\hspace{-0.5pt}2}}}(\omega)$ containing two monopole modes, obtained from numerical NMA for a Yukawa ball with $N=16$ particles and a screening strength of $\kappa=4.42\,{d_c}^{-1}$. The dashed lines indicate the position of the poles of equation (\ref{eq:spectrum}). b) In case of a vanishing BM no eigenvector is parallel to ${\vecn r}^*$, but generally many eigenvectors, e.g., ${\vecn r}_j,{\vecn r}_k,{\vecn r}_l$, have a non-vanishing projection on it giving rise to a (differnt) monopole oscillation. }
\label{fig:spectrum}\label{fig:BM_in_3N_space}
\end{figure}

In the case that a system has a BM, the MO describes the same motion as the BM and thus has the same frequency. However, in the opposite case, as for most Yukawa systems, there are several monopole oscillations with different frequencies. This property is directly apparent in the spectrum $\Phi_{\dot{R^{\hspace{-0.5pt}2}}}(\omega)$ of the mean square radius motion
\begin{equation}
	\dot{R^{\hspace{-0.5pt}2}}(t)=2N^{-1}\sum_{j=1}^{3N} \dot{c}_j(t){\vecn r}_j\cdot{\vecn r}^*+2N^{-1}\sum_{j=1}^{3N} c_j(t)\dot{c}_j(t),
\end{equation}
which may be derived by the square of its Fourier transform $\Phi_{\dot{R^{\hspace{-0.5pt}2}}}(\omega)=|\mathcal{F}_{\dot{R^{\hspace{-0.5pt}2}}}(\omega)|^2$. For the mentioned excitation the normal coordinates are given by $c_j(t\geq0)=\xi{\vecn r}^*\cdot{\vecn r}_j\cos(\omega_j t)$ and $c_j(t<0)=\xi{\vecn r}^*\cdot{\vecn r}_j$. Thus, for small amplitudes ($\xi\ll1$) the spectrum is given by
\begin{equation}\label{eq:spectrum}
	\Phi_{\dot{R^{\hspace{-0.5pt}2}}}(\omega)=\frac{2\xi^2}{\pi N^2}\sum_{j,k=1}^{3N}\frac{\bigl({\vecn r}_j\cdot{\vecn r}^*\bigr)^2\omega_j^2\bigl({\vecn r}_k\cdot{\vecn r}^*\bigr)^2\omega_k^2}{\bigl(\omega^2-\omega_j^2\bigr)\bigl(\omega^2-\omega_k^2\bigr)}.
\end{equation}
If the BM exists, one eigenvector ${\vecn r}_j$ is given by ${\vecn r}_{\rm BM}$, which is parallel to ${\vecn r}^*$, whereas all other eigenvectors are orthogonal to ${\vecn r}^*$. Hence the spectrum contains only a single peak at the breathing frequency. In contrast, if no BM exists, generally many eigenvectors ${\vecn r}_j$ will have a non-vanishing projection on ${\vecn r}^*$ [cf. figure \ref{fig:BM_in_3N_space} b)], so that the spectrum contains several MO with different frequencies $\omega_{\rm MO}=\omega_j$ and amplitudes.

Within figure \ref{fig:spectrum} a) such a spectrum is shown for a Yukawa ball with $N=16$ particles and a screening strength of $\kappa=4.42\,{d_c}^{-1}$. One clearly sees two peaks, thus this system contains two different monopole oscillations. The corresponding frequencies are $\omega_{\rm MO}^{(1)}=2.34905 \omega_0$ and $\omega_{\rm MO}^{(1)}=2.35954 \omega_0$.

\section{Experimental results}\label{experimental_results}\label{ex_s}
A very intriguing experimental realization of harmonically confined (screened) Coulomb systems are the so-called Yukawa balls in dusty plasmas. There, a finite number of plastic microspheres are dropped into the plasma of a parallel plate rf discharge. In the plasma, the particles arrange in concentric spherical shells (see e.g. \cite{Arp04,Block08,Kaeding08}) which is expected to be the ground state configuration of harmonically trapped 3D charged-particle systems \cite{Hasse91,Ludwig05,Apolinario07}.

The harmonic confinement in the plasma is realized by a combination of various plasma forces on the microspheres. In the vertical direction, the action of gravity is compensated by an upward force due to the electric field in the space-charge sheath above the lower electrode and an upward thermophoretic force due to heating of the lower electrode. Horizontally, the particles are confined by a glass cube placed onto the electrode. This results in a 3D harmonic confinement of the charged microspheres (see e.g. \cite{Arp05} for details). The charge on the typically used microspheres of
$3.5~\mu$m diameter is of the order of $10^3$ elementary charges, the interparticle distance is of about $500~\mu$m, and the shielding length in the plasma is of the same order as the interparticle distance. Thus, screening strengths of $\kappa\approx 1$ are typical in these systems \cite{bonitz_strutural_2006,Block08,Kaeding08} justifying the term ``Yukawa balls'' for our systems.

In the experiment, the normal modes of Yukawa balls are derived from the thermal Brownian motion of the particles around their equilibrium positions. The thermal particle motion is recorded with a stereoscopic video camera setup and the full 3D dynamics is reconstructed \cite{Kaeding07,Kaeding08}. Experimentally, the normal mode spectra are obtained as the spectral power density $S_{\ell} (\omega)$ from the Fourier transform for the mode number $\ell=1\ldots3N$ by \cite{Melzer03}
\begin{equation}\label{eq:Sl}
	S_{\ell}(\omega) = \frac{2}{T}\left|\int\limits_0^T \dot{c}_{\ell} (t)
	e^{-\rmi\omega t}\rmd t\right|^2 \quad.
\end{equation}
Here, $\dot{c}_{\ell}(t) = \dot{\vecn r}(t)\cdot{\vecn r}_{\ell}$ are the projections of the Brownian particle fluctuations with velocity $\dot{\vecn r}(t)$ onto the eigenvector of mode $\ell$, i.e., the normal mode oscillation pattern of this mode (see e.g. Ref.~\cite{Melzer03}).

\begin{figure}
	\centering
	\includegraphics[width=0.8\textwidth,clip=true]{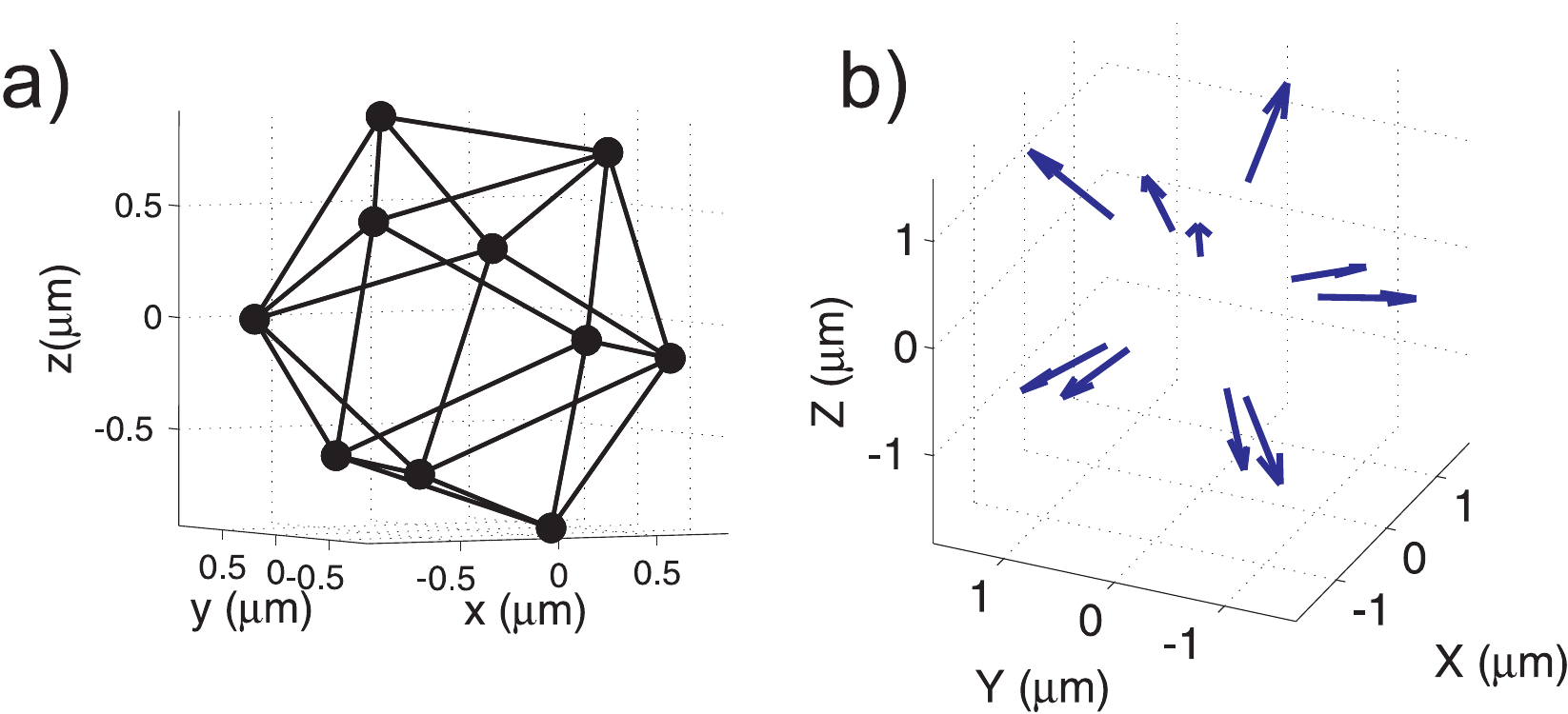}
	\caption{a) Bond structure model of a 3D dust cluster (Yukawa ball) with $N=10$ particles. Note the clear spherical structure. b) Breathing mode for this Yukawa ball derived from the experiment.}
	\label{fig:struct_10}
\end{figure}

The experimental mode spectrum is shown below for a Yukawa ball with $N=10$ particles. In equilibrium, the particles arrange on a single spherical shell as shown in Fig.~\ref{fig:struct_10}. As an example, also the breathing mode obtained from the experimental particle positions is presented. The 3D coherent radial particle motion is clearly seen (In our Yukawa balls, the deviation from a pure Coulomb interaction is not that strong, thus the breathing mode is not much distorted here).

\begin{figure}
	\centering
	\includegraphics[width=0.5\textwidth,clip=true]{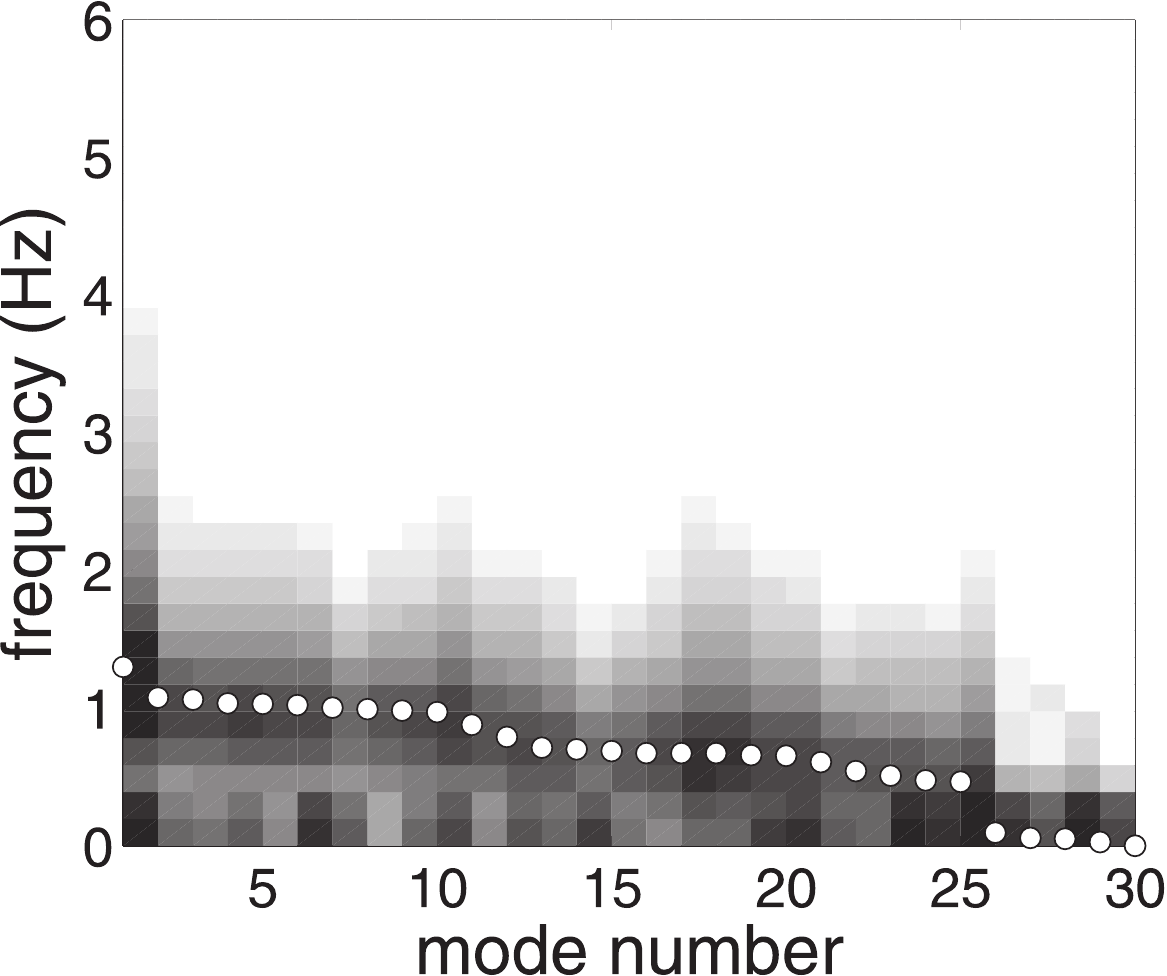}
	\caption{Experimental normal mode spectrum for a Yukawa ball with $N=10$. Here, the spectral power density is shown in gray scale (dark colors correspond to high power densities). The white dots indicate the best fit of the theoretical normal mode frequencies to the spectral power density.}
	\label{fig:spectrum_exp}
\end{figure}

The experimental normal mode spectrum $S_\ell (\omega)$ is shown in Fig.~\ref{fig:spectrum_exp}. It is seen that the mode frequencies are concentrated in a narrow band at very low frequencies of about 1-2~Hz. The dynamics of Yukawa balls occurs, due to the high mass of the microspheres, on very long time scales enabling the observation with video cameras. Also, the mode spectrum closely follows the expected mode frequencies from the mode analysis. The theoretical mode frequencies effectively depend only on the strength of the confinement $\omega_0$ in Eq.~(\ref{eq:hamiltonian}). By fitting the expected eigenfrequencies to the observed spectral power density a nice agreement is obtained and the confinement frequency is obtained. From the physical dimensions of the Yukawa ball and $\omega_0$ the dust charge is determined as $Z\approx 900$ for our 3.47~$\mu$m particles \cite{Ivanov08}.

\section{Normal modes in the presence of dissipation}
\begin{figure}
	\begin{center}
		\includegraphics[height=5cm,clip=true]{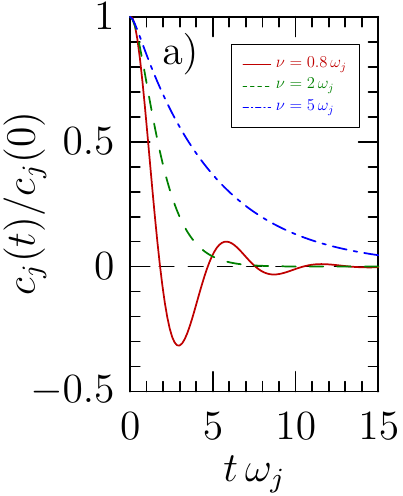}
		\includegraphics[height=5cm,clip=true]{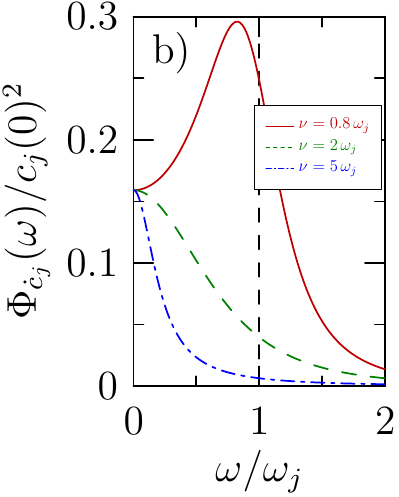}
		\includegraphics[height=5cm,clip=true]{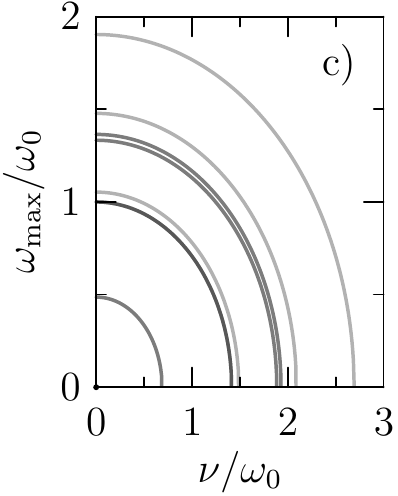}
	\end{center}
	\caption{(color online) a) Illustration of the damped dynamics of a normal mode $c_j(t)$ for three values of $\nu$ with the condition $\dot{c}_j(t<0)=0$. b) The corresponding spectra $\Phi_{\dot{c}_j}(\omega)$. The vertical dashed line indicates the peak position in the undamped case. c) Damping dependence of all eigenfrequencies ($\omega_{\rm max}$ denotes the peak position of the spectrum) of a Yukawa Ball with $N=5$ and $\kappa=1\,{d_c}^{-1}$. Darker lines highlight degenerate normal modes.}
\label{fig:mode_damping}
\end{figure}

The above presented theoretical investigations of normal modes neglect several effects present in the experiment -- most of all finite-temperature and dissipation. While inclusion of the former exceeds plain mechanics and requires statistics the latter can be taken into account by extending the equations of motion (\ref{eq:newton}) by a friction force $-m\nu\dot{{\vecn r}}$ with $\nu>0$, and we obtain
\begin{equation}
		{\vec 0}=m\ddot{{\vecn r}}+m\nu\dot{{\vecn r}}+\bnabla U({\vecn r})=m\sum_{j=1}^{3N}\left[\ddot{c}_j(t)+\nu\dot{c}_j(t)+\lambda_j c_j(t)\right]{\vecn r}_j.
\end{equation}
Hence, each normal coordinate fulfills $0=\ddot{c}_j(t)+\nu\dot{c}_j(t)+\lambda_j c_j(t)$ analogous to the single damped harmonic oscillator, and now depends on the damping strength $\nu$. The solution reads ($\omega_j=\sqrt{\lambda_j}$) \cite{hanno_dip}
\begin{equation}
	c_j(t)=A_j\begin{cases}
		\rme^{-\nu t/2}\cos\left(\sqrt{\omega_j^2-\nu^2/4}\;t+B_j\right)& \text{ for } \nu < 2\omega_j\\
		\rme^{-\nu t/2}\left[1+B_j t\right]& \text{ for } \nu = 2\omega_j\\
		\rme^{-\nu t/2}\left[\rme^{\sqrt{\nu^2/4-\omega_j^2}\;t}+B_j\rme^{-\sqrt{\nu^2/4-\omega_j^2}\;t}\right]& \text{ for } \nu > 2\omega_j
	\end{cases}
\end{equation}
Again, the constants $A_j$ and $B_j$ have to be determined from the initial conditions ${\vecn r}(0),\dot{\vecn r}(0)$ of the excitation. The three cases correspond to normal modes, which are underdamped, critically damped, and overdamped, respectively. Within figure \ref{fig:mode_damping} a) the dynamics of one mode $j$ are illustrated for these three cases for the condition $\dot{\vecn r}_j(0)=0$.

The modifications due to damping can nicely be seen within the spectrum of one normal mode, which is given in case of the aforementioned excitation by
\begin{equation}
	\Phi_{\dot{c}_j}(\omega)=\frac{\xi^2\bigl({\vecn r}^*\cdot{\vecn r}_j\bigr)^2\omega_j^4}{2\pi\bigl(\nu^2\omega^2+(\omega^2-\omega_j^2)^2\bigr)}.
\label{phi}
\end{equation}
In the undamped case ($\nu=0$) this spectrum just contains a single distinct peak centered at the normal mode frequency. In contrast, dissipation causes a red-shift as well as a broadening of this peak. These effects are shown in figure \ref{fig:mode_damping} b). The frequency shift of the peak at $\omega_{\rm max}$ in dependence on damping is displayed in figure \ref{fig:mode_damping} c) in more detail for the example of a Yukawa Ball with $N=5$ particles.

\begin{figure}
	\begin{center}
		\includegraphics[width=6cm,clip=true]{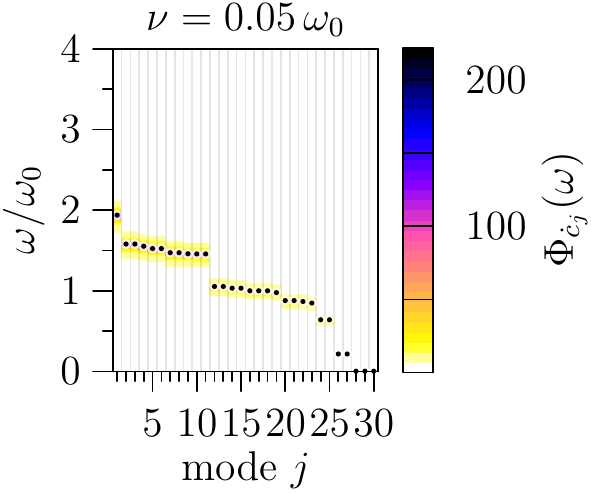}
		\includegraphics[width=6cm,clip=true]{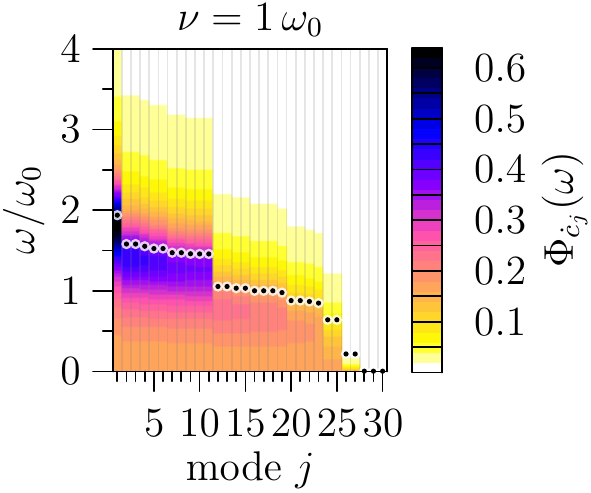}
		\includegraphics[width=6cm,clip=true]{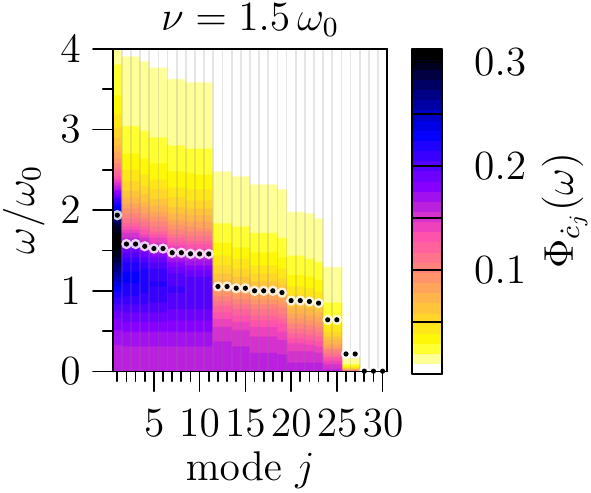}
		\includegraphics[width=6cm,clip=true]{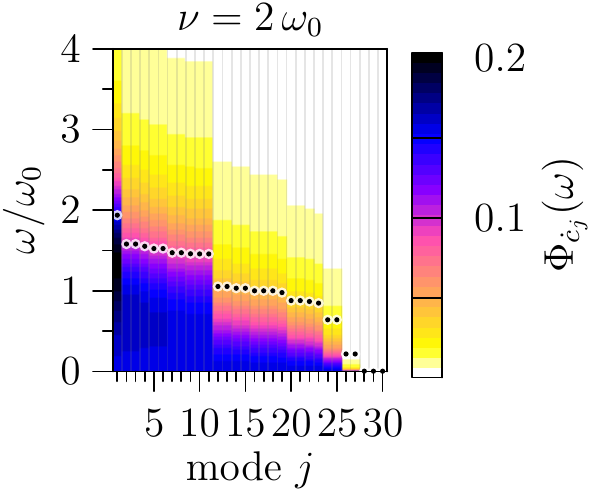}
	\end{center}
	\caption{(color online) Theoretical normal mode spectrum (\ref{phi}) for the cluster $N=10$ and $\kappa d_c=1$ for four values of the damping $\nu/\omega_0$ indicated in the figure. As initial conditions we chose $c_j(t=0)=c_0$ for all modes. 
For comparison, the undamped spectum is shown in all figures by black dots.}
\label{fig:spectrum_damping}
\end{figure}
Finally, in Fig.~\ref{fig:spectrum_damping} we present the complete power spectrum for different damping parameters, for the example $N=10$. Again, the red shift and peak broadening are evident. At sufficiently large damping more and more normal modes vanish, and an increasing fraction of the power density is concentrated around $\omega=0$. 
These theoretical results show a very similar behavior as the experimental data of Fig.~\ref{fig:spectrum_exp} and allow for a sensitive diagnostics of relevant experimental parameters, including the particle charge, cf. Sec.~\ref{ex_s}. A detailed theory-experiment comparison will be presented in a forthcoming paper.
\section{Summary and outlook}
Within this work we presented a theoretical and experimental analysis of the normal modes of three-dimensional spherially confined Yukawa clusters. Based on the predictions of Ref.~\cite{henning_existence_2008}, we analyzed the breathing mode in more detail, showed the close relation to the existence of multiple monopole oscillations in Yukawa systems and presented new numerical results. Finally, we investigated the influence of dissipation on the mode spectrum.
\section*{Acknowledgements}
This work is supported by the Deutsche Forschungsgemeinschaft via SFB-TR 24, projects A3, A5 and A7.

\end{document}